# Residual Entropy of a Two-dimensional Ising Model with Crossing and Four-spin Interactions

De-Zhang Li[1], Yu-Jun Zhao[1], Yao Yao[1] and Xiao-Bao Yang[1*]

[1] Department of Physics, South China University of Technology, Guangzhou 510640, China.

[*] Corresponding authors. Correspondence and requests for materials should be addressed to X.-B. Y. (email: scxbyang@scut.edu.cn).

# Abstract

We study the residual entropy of a two-dimensional Ising model with crossing and four-spin interactions, both for the case that in zero magnetic field and that in an imaginary magnetic field $i(\pi/2)k_B T$. The spin configurations of this Ising model can be mapped into the hydrogen configurations of square ice with the defined standard direction of the hydrogen bonds. Making use of the equivalence of this Ising system with the exactly solved eight-vertex model and taking the low temperature limit, we obtain the residual entropy. Two soluble cases in zero field and one soluble case in imaginary field are examined. In the case that the free-fermion condition holds in zero field, we find the ground states in low temperature limit include the configurations disobeying the ice rules. In another case in zero field that the four-spin interactions are $-\infty$, and the case in imaginary field that the four-spin interactions are 0, the residual entropy exactly agrees with the result of square ice determined by Lieb in 1967. In the solutions of the latter two cases, we have shown alternative approaches to the residual entropy problem of square ice.

## I. Introduction

Exact solution of Ising model formulated on certain lattice is of interest in statistical physics for a long time. The simple one-dimensional Ising model was solved by Ising himself early in 1920s[1]. In 1944, Onsager published his famous derivation of the solution for the two-dimensional Ising model with nearest-neighbour interactions and without external field[2]. The result of Onsager was obtained by transfer matrix method, and rederived from various approaches[3-5]. In 1952, Lee and Yang obtained a solution for the two-dimensional Ising model with nearest-neighbour interactions in an imaginary external field[6]. This solution has also been determined from a variety of different approaches[7-12]. While the Ising models with nearest-neighbour interactions on a square lattice are exactly solved, those formulated on a checkerboard lattice are more difficult to treat[13, 14]. This is caused by the frustrated structure of the checkerboard lattice.

Residual entropy, determined directly by the ground state degeneracy, exists in many frustrated systems such as frustrated Ising model[15, 16] and ice system[17]. Early in 1930s the theoretical explanation of the residual entropy of ice was proposed by the ice rules[18, 19], which states that in the ice lattice there is only one hydrogen between every pair of nearest-neighbour oxygens to form a hydrogen bond and there are two hydrogens adjacent to each oxygen to constitute a $H_2O$ molecule. Denoting $W$ as the number of hydrogen configurations obeying the ice rules and $N_{H_2O}$ as the number of $H_2O$ molecules, the residual entropy can be expressed as $S/k_B = \frac{1}{N_{H_2O}} \ln W$. Pauling first made a rough estimate for the residual entropy by mean field approximation[19]. This estimate was found to be a lower bound by Onsager and Dupuis[20]. Nagle used a series expansion method developed from DiMarzio and Stillinger's approach[21] and obtained an advanced theoretical approximation[22]. Nagle's result is in excellent agreement with experiment[23] and usually treated as the best theoretical estimate for the three-dimensional ice so far. In 1967, the exact solution for the residual entropy of square ice

$$S/k_B = \frac{3}{2}\ln\left(\frac{4}{3}\right) \qquad (1)$$

was published by Lieb using the transfer matrix method[24, 25], which directly evaluated the number of hydrogen configurations obeying the ice rules on square lattice to solve the problem. In the context of Ising models, the residual entropy is also well appreciated at least since the exact solutions of the triangular model[26] and the Kagomé model[27]. Anderson first showed the direct connection between the residual entropy of ice system and that of Ising model[28]. Especially, the close relation of square ice with the two-dimensional Ising model has been discussed by Lieb and Wu[29], and by Liebmann[15]. Lieb and Wu[29] further showed the mapping of some special cases of two-dimensional Ising system into the eight-vertex model[30], and into the sixteen-vertex model like the general F model[31] and the general KDP model[32]. Accurate calculation of the residual entropy of frustrated systems remains a challenging task, even for the two-dimensional checkerboard Ising model.

In this article, our focus is the study of a two-dimensional Ising model on a checkerboard lattice with crossing and four-spin interactions. In Section II, we introduce the Ising model and verify the connection between this model and square ice. In Section III, following the work of Refs. 13 and 14, we examine two soluble cases in zero magnetic field and one soluble case in an imaginary magnetic field. For all these cases, we find the approach to the residual entropy by taking the low temperature limit of the partition function, instead of using the transfer matrix method or the combinational methods. The difference between the results in these cases is demonstrated, and the relation of the result with the residual entropy of square ice [Eq. (1)] in each case is discussed. The conclusions are outlined in Section IV.

**II. Model**

Consider a two-dimensional Ising model of $N$ spins with crossing and four-spin interactions on a checkerboard lattice as shown in Fig. 1. The interaction energy of each crossed square is

$$E(s_1,s_2,s_3,s_4) = J(s_1s_2 + s_2s_3 + s_3s_4 + s_4s_1 + s_1s_3 + s_2s_4) + \Delta s_1s_2s_3s_4, \quad (2)$$

where $J$ is the two-spin interaction and $\Delta$ is the four-spin interaction. In our case, $J$ is a positive constant. In the presence of an external magnetic field, the total Hamiltonian of the Ising model is

$$H(\{s_i\}) = \sum_{\text{crossed square}} E(s_1,s_2,s_3,s_4) - H_{ex}\sum_{i=1}^{N} s_i, \quad (3)$$

where the periodic boundary condition is taken into account. The partition function can then be written as

$$Z_N = \sum_{s_i=\pm 1} \exp[-\beta H(\{s_i\})] \quad (4)$$

with $\beta = 1/k_B T$ and $k_B$ is the Boltzmann constant. The energy levels of $E(s_1,s_2,s_3,s_4)$ within each crossed square and the corresponding degeneracy and spin configurations are listed in Table I. Each crossed square at a certain energy level is denoted as type A, B or C as shown in the table and one type A crossed square will be expressed as 1(A) hereafter.

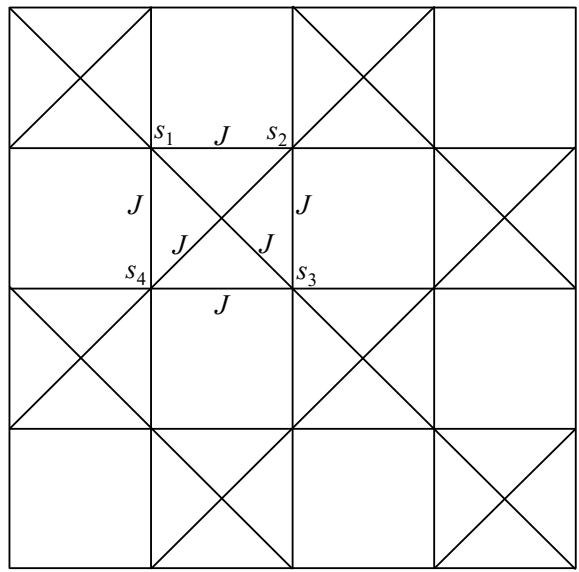

**Fig. 1**. The two-dimensional Ising model with crossing interactions (the four-spin interactions are not shown).

**Table I. The energy levels and the corresponding degeneracy and spin configurations of each crossed square.**

| $E$ | $-2J+\Delta$ | $-\Delta$ | $6J+\Delta$ |
|---|---|---|---|
| $g(E)$ | 6 | 8 | 2 |
| Spin configurations | $2\times(+1)+2\times(-1)$ | $3\times(+1)+1\times(-1)$ or $3\times(-1)+1\times(+1)$ | $4\times(+1)$ or $4\times(-1)$ |
| Type | A | B | C |

To verify the connection between this Ising model and square ice, we should first define the standard direction (denoted as +1) of the hydrogen bonds in square ice as shown in Fig. 2. One sees that each crossed square surrounded by four spins can be mapped into an oxygen lattice point of square ice, and the value +1/-1 of the four spins corresponds to the direction of the four hydrogen bonds around this site respectively. Therefore the hydrogen configurations of square ice can be mapped into the spin configurations of the Ising model, and those obeying the ice rules correspond to the configurations with two +1 spins and two -1 spins in every crossed square, i.e., $(N/2)$ (A). Note that $N/2$ is the number of crossed squares and is also the number of H$_2$O molecules. Then the ground state degeneracy $g(E_0)$ of the Ising model in the large-$N$ limit directly leads to the residual entropy

$$S/k_B = \lim_{N\to\infty} \frac{1}{N/2} \ln\left[g(E_0)\right]. \tag{5}$$

In the low temperature limit only the ground states with the Boltzmann weight $e^{-\beta E_0}$ will appear in the partition function $Z_N$, i.e., $Z_N \simeq g(E_0)e^{-\beta E_0}$. Thus we can obtain the expression for the residual entropy

$$S/k_B = \lim_{\beta \to \infty}\left\{\lim_{N \to \infty}\frac{1}{N/2}\left(\ln Z_N + \beta E_0\right)\right\} . \qquad (6)$$

The accurate calculation of the residual entropy defined above requires the exact solution of the partition function $Z_N$ and the value of ground state energy $E_0$.

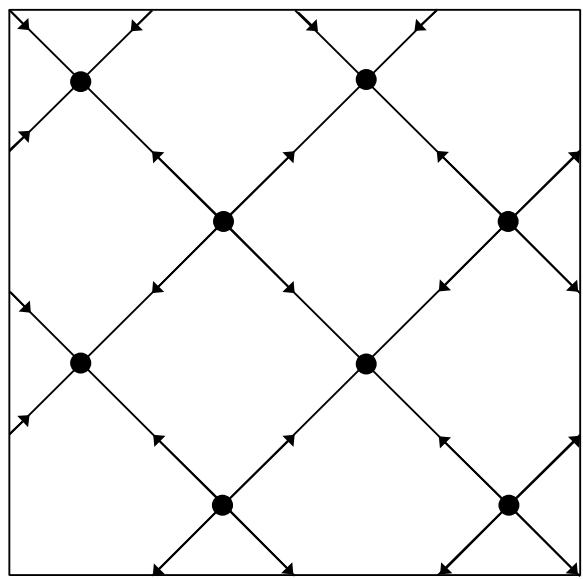

**Fig. 2**. The standard direction (+1) of the hydrogen bonds in square ice.

### III. Results and Discussions

#### A. Zero Field $H_{ex} = 0$

To evaluate the partition function $Z_N$ in the case of zero field $H_{ex} = 0$, Giacomini introduced a mapping of the Ising model into an eight-vertex model with $N/2$ vertices and showed the equivalence[13]

$$Z_N = \frac{1}{2^{N/2}}\exp\left[-\beta\frac{N}{2}(6J+\Delta)\right]\times Z_{8v} . \qquad (7)$$

Following the mapping procedure[13], the vertex weights of the eight-vertex model in our case can be obtained

$$\omega_1 = 1 + 4e^{\beta(6J+2\Delta)} + 3e^{8\beta J},$$
$$\omega_2 = 1 - 4e^{\beta(6J+2\Delta)} + 3e^{8\beta J}, \quad (8)$$
$$\omega_3 = \omega_4 = \omega_5 = \omega_6 = \omega_7 = \omega_8 = 1 - e^{8\beta J}.$$

Two soluble cases of this eight-vertex model are considered in Ref. 13, namely, the free-fermion case[33, 34] and the $\Delta = -\infty$ case. In both cases, the explicit solution of $Z_N$ can be obtained. We will present the solution and analyse the low temperature limit in both cases respectively.

### A.1. Free-fermion Model

The eight-vertex model can be exactly solved when the free-fermion condition[33, 34]

$$\omega_1\omega_2 + \omega_3\omega_4 = \omega_5\omega_6 + \omega_7\omega_8 \quad (9)$$

is satisfied. Substituting Eq. (8) into Eq. (9), the free-fermion condition is expressed as

$$\exp(4\beta\Delta) = \cosh(4\beta J) . \quad (10)$$

Obviously, the four-spin interaction $\Delta$ is temperature-dependent when $J$ is a positive constant. It is trivial to demonstrate that at finite temperature $0 < \Delta < J$ and in the low temperature limit ($\beta \to \infty$) $\Delta \to J$. Then we can see from Table I that at zero temperature the energy of $2\times(+1)+2\times(-1)$ configurations within a crossed square equals to that of $3\times(+1)+1\times(-1)$ / $3\times(-1)+1\times(+1)$ configurations. Therefore, the ground states of the system at finite temperature are exactly the configurations with $(N/2)$ (A), but at zero temperature they include the configurations with (B). That is, in the low temperature limit the ground states include the configurations disobeying the ice rules. The exact expression for the partition function of the free-fermion model is given by[33, 34]

$$\lim_{N\to\infty}\frac{1}{N/2}\ln Z_{8v}=\frac{1}{8\pi^2}\int_0^{2\pi}d\theta\int_0^{2\pi}d\phi\ln\left[2a+2b\cos\theta+2c\cos\phi+2d\cos(\theta+\phi)+2e\cos(\theta-\phi)\right] \quad (11)$$

with

$$\begin{aligned}2a &= \omega_1^2+\omega_2^2+\omega_3^2+\omega_4^2,\\ b &= \omega_1\omega_3-\omega_2\omega_4,\\ c &= \omega_1\omega_4-\omega_2\omega_3,\\ d &= \omega_3\omega_4-\omega_7\omega_8,\\ e &= \omega_3\omega_4-\omega_5\omega_6.\end{aligned} \quad (12)$$

Using the equivalence between the Ising system and the eight-vertex model [Eq. (7)] and simply substituting Eq. (8) and Eq. (10) into Eqs. (11)-(12), the partition function of the Ising system can then be written as

$$\lim_{N\to\infty}\frac{1}{N/2}\ln Z_N = \beta(2J-\Delta)+\frac{1}{8\pi^2}\int_0^{2\pi}d\theta\int_0^{2\pi}d\phi\ln\left[\left(3+e^{-8\beta J}\right)^2-2\sqrt{2}\left(1-e^{-8\beta J}\right)\sqrt{1+e^{-8\beta J}}\left(\cos\theta+\cos\phi\right)\right]. \quad (13)$$

It is easy to verify that the ground state energy of the Ising system at finite temperature is $E_0=\frac{N}{2}(-2J+\Delta)$. Taking $Z_N$ in Eq. (13) into account, the right-hand side of Eq. (6) at finite temperature becomes

$$\lim_{N\to\infty}\frac{1}{N/2}(\ln Z_N+\beta E_0)=\frac{1}{8\pi^2}\int_0^{2\pi}d\theta\int_0^{2\pi}d\phi\ln\left[\left(3+e^{-8\beta J}\right)^2-2\sqrt{2}\left(1-e^{-8\beta J}\right)\sqrt{1+e^{-8\beta J}}\left(\cos\theta+\cos\phi\right)\right]. \quad (14)$$

Now we take $\beta\to\infty$ to obtain the entropy in the low temperature limit

$$\begin{aligned}S/k_B &= \frac{1}{8\pi^2}\int_0^{2\pi}d\theta\int_0^{2\pi}d\phi\ln\left[9-2\sqrt{2}(\cos\theta+\cos\phi)\right]\\ &\simeq 1.07052.\end{aligned} \quad (15)$$

This value is significantly larger than the residual entropy of square ice [Eq. (1)]. As mentioned before, the ground states at zero temperature include the configurations with (B), which disobey the ice rules and result in a larger configurational entropy. To verify the behaviour of the Ising system in the low temperature limit, we express the partition function in the form of the summation of all energy levels in the system

$$Z_N = e^{-\beta E_0}\left[g(E_0) + \sum_{j=1}g(E_j)e^{-\beta \delta E_j}\right]. \qquad (16)$$

Here $\delta E_j$ is the energy difference between each energy level and the ground state, i.e., $\delta E_j = E_j - E_0$ and $g(E_j)$ is the degeneracy of the energy level. At finite temperature, the order of the energy within three types of crossed square is $(A) < (B) < (C)$. Therefore, $g(E_j)$ depends only on the configurations consist of (A), (B) and (C), but not on the temperature. E.g., the first excited states at finite temperature are the configurations with $(N/2-2)(A) + 2(B)$ so that $g(E_1)$ is just the number of all possible combinations of $(N/2-2)(A)$ and $2(B)$ on the lattice shown in Fig. 1. Since $g(E_j)$ is temperature-independent, it is straightforward to express the residual entropy in Eq. (6) using Eq. (16)

$$S/k_B = \lim_{N\to\infty}\frac{1}{N/2}\ln\left[g(E_0) + \sum_{j=1}g(E_j)\lim_{\beta\to\infty}e^{-\beta \delta E_j}\right]. \qquad (17)$$

As the ground states at finite temperature are the configurations with $(N/2)(A)$, $\delta E_j$ for a certain energy level just depends on the number of (B) and (C). For the configuration $(N/2-x_j-y_j)(A) + x_j(B) + y_j(C)$ at the energy level $j$, $\delta E_j = 2x_j(J-\Delta) + 8y_jJ$. If $y_j > 0$, we have $\lim_{\beta\to\infty}e^{-\beta \delta E_j} = 0$. Then the energy levels corresponding to the configurations with (C) have no contribution to the summation in Eq. (17), and we may consider those corresponding to $(N/2-x_j)(A) + x_j(B)$ only. Notice that $x_j$ should be even for these configurations, i.e., $x_j = 2z_j$. E.g., for the first excited state at finite temperature $x_1 = 2$ and $z_1 = 1$. Using the free-fermion condition Eq. (10), we have

$$\lim_{\beta \to \infty} e^{-\beta \delta E_j} = \left(\frac{1}{2}\right)^{z_j}. \tag{18}$$

Substituting this factor into the expression of residual entropy produces

$$S/k_B = \lim_{N \to \infty} \frac{1}{N/2} \ln\left[g(E_0) + \sum_{j=1}^{*} g(E_j)\left(\frac{1}{2}\right)^{z_j}\right], \tag{19}$$

where the summation $\sum^{*}$ is taken over all the energy levels corresponding to the configurations without (C). As shown in Eq. (15), we obtain the low temperature limit for this free-fermion model

$$\lim_{N \to \infty} \frac{1}{N/2} \ln\left[g(E_0) + \sum_{j=1}^{*} g(E_j)\left(\frac{1}{2}\right)^{z_j}\right] = \frac{1}{8\pi^2} \int_0^{2\pi} d\theta \int_0^{2\pi} d\phi \ln\left[9 - 2\sqrt{2}(\cos\theta + \cos\phi)\right] \tag{20}$$
$$\simeq 1.07052 .$$

Compared with $\lim_{N \to \infty} \frac{1}{N/2} \ln g(E_0)$, which is the residual entropy of square ice, the result in this case is larger because the ground states at zero temperature include the configurations with (B) that disobey the ice rules.

### A.2. $\Delta = -\infty$

Another soluble case of the eight-vertex model is that the condition

$$\omega_1 = \omega_2, \quad \omega_3 = \omega_4 \tag{21}$$

is satisfied[35, 36]. One can see from the vertex weights in Eq. (8) that in this case the four-spin interaction $\Delta$ should be $-\infty$. To study the exact thermodynamic properties of the system in this case, we should first define the effective Hamiltonian $\bar{H}$ and the corresponding effective partition function $\bar{Z}_N$. The energy within each crossed square is modified as $\bar{E} = E - \Delta$ with $E$ defined in Eq. (2). Then the effective Hamiltonian $\bar{H}$ is simply $\bar{H} = H - \frac{N}{2}\Delta$, which leads to the effective partition function $\bar{Z}_N = Z_N / \exp\left[-\beta \frac{N}{2}\Delta\right]$. From Table II we can easily show that the order of the modified energy of the three types of crossed square is $(A) < (C) < (B) = +\infty$. Obviously, only the configurations without (B) have contributed to the partition function $\bar{Z}_N$. The ground states are exactly the configurations with $(N/2)$ (A) obeying the ice rules. Therefore, the residual entropy of this system exactly agrees with that of square ice. Making use of the equivalence with the eight-vertex model shown in Eq. (7), the exact solution of $\bar{Z}_N$ can be obtained

$$\bar{Z}_N = \frac{1}{2^{N/2}} \exp[-3\beta NJ] \times Z_{8v} . \tag{22}$$

Here the vertex weights are

$$\begin{aligned} \omega_1 &= \omega_2 = 1 + 3e^{8\beta J}, \\ \omega_3 &= \omega_4 = \omega_5 = \omega_6 = \omega_7 = \omega_8 = 1 - e^{8\beta J}. \end{aligned} \tag{23}$$

Notice that the ground state energy of the system is $\bar{E}_0 = -NJ$. Similar to Eq. (6), the expression for the residual entropy in this case is given by

$$\begin{aligned} S/k_B &= \lim_{\beta \to \infty} \left\{ \lim_{N \to \infty} \frac{1}{N/2} \left( \ln \bar{Z}_N + \beta \bar{E}_0 \right) \right\} \\ &= \lim_{\beta \to \infty} \left\{ \lim_{N \to \infty} \frac{1}{N/2} \ln Z_{8v} - \ln 2 - 8\beta J \right\}. \end{aligned} \tag{24}$$

**Table II. The modified energy levels and the corresponding degeneracy and spin configurations of each crossed square.**

| $\bar{E}$ | $-2J$ | $-2\Delta$ | $6J$ |
|---|---|---|---|
| $g(\bar{E})$ | 6 | 8 | 2 |

| Spin configurations | $2\times(+1)+2\times(-1)$ | $3\times(+1)+1\times(-1)$ or $3\times(-1)+1\times(+1)$ | $4\times(+1)$ or $4\times(-1)$ |
|---|---|---|---|
| Type | A | B | C |

The eight-vertex model is solved by Baxter[35-38] when the condition Eq. (21) holds. Following Baxter's work, the partition function $Z_{8v}$ can be expressed as the function of four quantities $w_j\,(j=1,\cdots,4)$ determined by the vertex weights in Eq. (23). Arranged in nonincreasing order $w_1 \geq w_2 \geq w_3 \geq w_4$, the values of $w_j\,(j=1,\cdots,4)$ are given by

$$\begin{aligned} w_1 &= 2e^{8\beta J}, \\ w_2 &= e^{8\beta J}+1, \\ w_3 &= e^{8\beta J}-1, \\ w_4 &= 0. \end{aligned} \quad (25)$$

To obtain the solution for $Z_{8v}$, we first determine the parameters $l$, $l'$, $\zeta$ and $V$ from $w_j\,(j=1,\cdots,4)$

$$\begin{aligned} l &= \sqrt{\frac{(w_1^2-w_4^2)(w_2^2-w_3^2)}{(w_1^2-w_3^2)(w_2^2-w_4^2)}}, \\ l' &= \sqrt{1-l^2}, \\ \operatorname{sn}(\zeta;l) &= \sqrt{\frac{w_1^2-w_3^2}{w_1^2-w_4^2}}, \\ \operatorname{sn}(V;l) &= \frac{w_4}{w_3}\operatorname{sn}(\zeta;l), \end{aligned} \quad (26)$$

where $\operatorname{sn}(\zeta;l)$ and $\operatorname{sn}(V;l)$ are Jacobian elliptic functions of modulus $l$. Next, we denote $K_l$ as the complete elliptic integral of the first kind of modulus $l$ and define

$$\mu = \frac{\pi\zeta}{K_l}, \quad U = \frac{\pi V}{K_l}, \quad q = \exp\left(-\frac{\pi K_{l'}}{K_l}\right). \quad (27)$$

Since our interest is the low temperature limit, we will make use of a more convenient expression for $Z_{8v}$ applicable in the low temperature regime instead of the original series form (Eq. (7.7) in Ref. 36). We see from Eq. (25) that in the low temperature regime $w_2 \to w_3$. According to Appendix E of Ref. 36, the partition function of this regime can be written as

$$\lim_{N \to \infty} \frac{1}{N/2} \ln Z_{8v} = \psi + h(0) - F_1 + F_2 \ . \tag{28}$$

The four terms $\psi$, $h(0)$, $F_1$ and $F_2$ in this expression are given by

$$\psi = \frac{1}{2} \ln \left[ \frac{\pi}{2l'^{1/2} K_l} \tan\left(\frac{\mu}{2}\right) \frac{\cos\left(\frac{U}{2}\right) + \cos\left(\frac{\mu}{2}\right)}{\cos\left(\frac{U}{2}\right) - \cos\left(\frac{\mu}{2}\right)} \right] + \frac{1}{8} \ln\left[ \left(w_1^2 - w_2^2\right)\left(w_1^2 - w_3^2\right)\left(w_2^2 - w_4^2\right)\left(w_3^2 - w_4^2\right) \right], \tag{29}$$

$$h(0) = \int_{-\infty}^{\infty} \frac{\sinh^2\left[(\pi - \mu)x\right]\left[\cosh(\mu x) - \cosh(Ux)\right]}{x \sinh(2\pi x) \cosh(\mu x)} dx \ , \tag{30}$$

$$F_1 = 2 \sum_{n=1}^{\infty} \frac{q^{2n}}{1 - q^{2n}} \frac{\sin^2(n\mu)\left[\cos(n\mu) - \cos(nU)\right]}{n \cos(n\mu)} \ , \tag{31}$$

and

$$F_2 = 4 \sum_{n=1}^{\infty} \frac{1}{2n-1} \frac{(-1)^n q^{(2n-1)\pi/\mu}}{1 - q^{(2n-1)\pi/\mu}} \cot\left[\frac{(2n-1)\pi^2}{2\mu}\right] \cos\left[\frac{(2n-1)\pi U}{2\mu}\right] , \tag{32}$$

respectively. Now it is straightforward to substitute Eq. (28) into Eq. (24) and obtain the expression for the residual entropy

$$S/k_B = \lim_{\beta \to \infty} \left\{ \left(\psi - \ln 2 - 8\beta J\right) + h(0) - F_1 + F_2 \right\} \tag{33}$$

with the four terms $\psi$, $h(0)$, $F_1$ and $F_2$ shown in Eqs. (29)-(32). Clearly, the residual entropy is exactly the summation of the low temperature limit of $\psi - \ln 2 - 8\beta J$, $h(0)$, $F_1$ and $F_2$.

Taking the low temperature limit $\beta \to \infty$ of $w_j (j = 1, \cdots, 4)$ in Eq. (25) we can easily find

$$l \to 0, \quad l' \to 1, \quad \zeta \to \frac{\pi}{3}, \quad V = 0, \tag{34}$$

$$K_l \to \frac{\pi}{2}, \quad K_{l'} \to +\infty, \quad \mu \to \frac{2\pi}{3}, \quad U = 0, \quad q \to 0 \tag{35}$$

and

$$\frac{1}{8}\ln\left[\left(w_1^2 - w_2^2\right)\left(w_1^2 - w_3^2\right)\left(w_2^2 - w_4^2\right)\left(w_3^2 - w_4^2\right)\right] - \ln 2 - 8\beta J \to \frac{1}{4}\ln\left(\frac{3}{16}\right). \tag{36}$$

The limit values of the four terms $\psi - \ln 2 - 8\beta J$, $h(0)$, $F_1$ and $F_2$ in Eq. (33) will be given respectively using Eqs. (34)-(36).

(1). $\psi - \ln 2 - 8\beta J$

It is straightforward to substitute Eqs. (34)-(36) into Eq. (29), and obtain

$$\lim_{\beta \to \infty}\left(\psi - \ln 2 - 8\beta J\right) = \ln\left(\frac{3}{2}\right). \tag{37}$$

(2). $h(0)$

Substituting Eq. (35) into Eq. (30) produces

$$\lim_{\beta \to \infty} h(0) = \int_{-\infty}^{\infty} \frac{\sinh^2\left(\frac{\pi}{3}x\right)\left[\cosh\left(\frac{2\pi}{3}x\right) - 1\right]}{x \sinh(2\pi x) \cosh\left(\frac{2\pi}{3}x\right)} dx. \tag{38}$$

Performing the variable transformation $y = \frac{2\pi}{3}x$ and inserting the identity $\frac{\cosh(y) - 1}{y} = \int_0^1 \sinh(yu)\, du$ into Eq. (38) gives

$$\lim_{\beta \to \infty} h(0) = \frac{1}{2}\int_0^1 du \int_{-\infty}^{\infty} dy \frac{\sinh(yu)\left[\cosh(y) - 1\right]}{\sinh(3y)\cosh(y)}. \tag{39}$$

The result of this integral obtained by MATHEMATICA is $\ln\left(\frac{16}{9\sqrt{3}}\right)$. That is,

$$\lim_{\beta \to \infty} h(0) = \ln\left(\frac{16}{9\sqrt{3}}\right). \tag{40}$$

(3). $F_1$

We can employ some series analysis techniques to deal with $F_1$. Starting from Eq. (31) with $\mu$, $U$ and $q$ in Eq. (35), we have

$$\lim_{\beta \to \infty} F_1 = 2\lim_{q \to 0} \sum_{n=1}^{\infty} \frac{q^{2n}}{1-q^{2n}} \frac{\sin^2\left(\frac{2n}{3}\pi\right)\left[\cos\left(\frac{2n}{3}\pi\right)-1\right]}{n\cos\left(\frac{2n}{3}\pi\right)}. \tag{41}$$

It is obviously to show

$$\lim_{\beta \to \infty} |F_1| < 2\lim_{q \to 0}\left\{\sum_{n=1}^{\infty} \frac{q^{2n}}{1-q^{2n}} \frac{2}{n \times \frac{1}{2}}\right\} = 8\lim_{q \to 0}\left\{\frac{q^2}{1-q^2} + \sum_{n=2}^{\infty} \frac{q^{2n}}{1-q^{2n}} \frac{1}{n}\right\}. \tag{42}$$

For the series term in the right-hand side of Eq. (42), one can see that

$$\sum_{n=2}^{\infty} \frac{q^{2n}}{1-q^{2n}} \frac{1}{n} < \int_1^{\infty} \frac{q^{2x}}{1-q^{2x}} \frac{1}{x} dx < \int_1^{\infty} \frac{q^{2x}}{1-q^{2x}} dx = \frac{\ln(1-q^2)}{\ln(q^2)}. \tag{43}$$

Then it is straightforward to find the right-hand side of Eq. (42) is 0. That is,

$$\lim_{\beta \to \infty} F_1 = 0. \tag{44}$$

(4). $F_2$

Using the similar techniques as in the analysis of $F_1$, we can conclude that

$$\lim_{\beta \to \infty} F_2 = 0. \tag{45}$$

Substituting the limit values of the four terms into Eq. (33), we obtain the result for the residual entropy

$$S/k_B = \ln\left(\frac{3}{2}\right) + \ln\left(\frac{16}{9\sqrt{3}}\right) = \frac{3}{2}\ln\left(\frac{4}{3}\right). \tag{46}$$

This result exactly agrees with that of square ice [Eq. (1)]. As mentioned before, the ground states of the system in this case are exactly the configurations obeying the ice rules. Then the residual entropy determined by the ground state degeneracy $\lim_{N\to\infty} \frac{1}{N/2} \ln g(\bar{E}_0)$ is consistent with the configurational entropy of square ice. In fact, we have shown an alternative derivation of this exact solution.

## B. Imaginary Field $H_{ex} = i(\pi/2)k_B T$

This case was first considered by Wu[14], inspired by the famous solution of the two-dimensional Ising model in an imaginary external field proposed by Lee and Yang[6]. Substitute Eq. (3) and the imaginary value of the magnetic field $H_{ex} = i(\pi/2)k_B T$ into Eq. (4), we may express the partition function as

$$Z_N = \sum_{s_i = \pm 1} i^N \prod_{i=1}^{N} s_i \times \exp\left[-\beta \sum_{\text{crossed square}} E(s_1, s_2, s_3, s_4)\right] \quad (47)$$

by using $i^{s_i} = i \times s_i$. Following Ref. 14, we assume $N$ to be multiples of 4 so that the factor $i^N$ in Eq. (47) equals to 1. The model in the presence of the imaginary field $H_{ex} = i(\pi/2)k_B T$ is solvable. Here we study the case that $\Delta = 0$. From Table I it is trivial to demonstrate that, the ground states in this case are exactly the configurations with $(N/2)$ (A). For the ground states, the number of -1 spins takes one half of $N$, which is also even such that $\prod_{i=1}^{N} s_i = 1$. With this consideration and realizing the ground state energy $E_0 = -NJ$, we can easily show that in the low temperature limit $Z_N$ has the same behaviour as $\bar{Z}_N$ in the case $\Delta = -\infty$ in zero field shown in Sec. III.A.2. Therefore the residual entropy in this case should be consistent with the result of square ice.

Wu introduced a mapping of this Ising model with $H_{ex} = i(\pi/2)k_B T$ into an exactly solved eight-vertex model with $N/2$ vertices and showed the exact equivalence[14]

$$Z_N = Z_{8v} . \qquad (48)$$

Following the mapping procedure[14, 33, 39], the vertex weights of the eight-vertex model in this case are given by

$$\begin{aligned}
\omega_1 &= \omega_2 = \omega_3 = \omega_4 = \omega_7 = \omega_8 = e^{-2\beta J} \sinh(-4\beta J), \\
\omega_5 &= \frac{1}{2}\left(3e^{2\beta J} - 4 + e^{-6\beta J}\right), \\
\omega_6 &= \frac{1}{2}\left(3e^{2\beta J} + 4 + e^{-6\beta J}\right).
\end{aligned} \qquad (49)$$

We can see that the condition Eq. (21) holds. As stated in Ref. 14, the partition function can be evaluated in this case if $\omega_5 \omega_6 > 0$. It is easy to verify that $\omega_5 > 0$ and $\omega_6 > 0$ so that this eight-vertex model is solvable. Similar to the solution in Sec. III.A.2, $Z_{8v}$ in this case can be expressed as the function of four quantities $w_j$ $(j=1,\cdots,4)$ given by

$$\begin{aligned}
w_1 &= \frac{e^{2\beta J}}{4}\left[\sqrt{\left(3 - 4e^{-2\beta J} + e^{-8\beta J}\right)\left(3 + 4e^{-2\beta J} + e^{-8\beta J}\right)} + \left(1 - e^{-8\beta J}\right)\right], \\
w_2 &= \frac{e^{2\beta J}}{2}\left(1 - e^{-8\beta J}\right), \\
w_3 &= \frac{e^{2\beta J}}{4}\left[\sqrt{\left(3 - 4e^{-2\beta J} + e^{-8\beta J}\right)\left(3 + 4e^{-2\beta J} + e^{-8\beta J}\right)} - \left(1 - e^{-8\beta J}\right)\right], \\
w_4 &= 0.
\end{aligned} \qquad (50)$$

In the low temperature regime it can be written as

$$\lim_{N\to\infty} \frac{1}{N/2} \ln Z_{8v} = \psi + h(0) - F_1 + F_2 , \qquad (51)$$

where the four terms $\psi$, $h(0)$, $F_1$ and $F_2$ are defined in Eqs. (29)-(32) and Eqs. (26)-(27) with $w_j$ $(j=1,\cdots,4)$ given in Eq. (50). Taking Eq. (51) and $E_0 = -NJ$ into account, the expression for the residual entropy in this case can be obtained directly from Eq. (6)

$$S/k_B = \lim_{\beta \to \infty}\{(\psi - 2\beta J) + h(0) - F_1 + F_2\} . \tag{52}$$

In the low temperature limit, Eqs. (34)-(35) still hold and we have

$$\frac{1}{8}\ln\left[(w_1^2 - w_2^2)(w_1^2 - w_3^2)(w_2^2 - w_4^2)(w_3^2 - w_4^2)\right] - 2\beta J \to \frac{1}{4}\ln\left(\frac{3}{16}\right) . \tag{53}$$

Then it is straightforward to demonstrate that

$$\lim_{\beta \to \infty}(\psi - 2\beta J) = \ln\left(\frac{3}{2}\right) \tag{54}$$

and the limit values of the other three terms $h(0)$, $F_1$ and $F_2$ are completely identical to the results in Eqs. (40), (44) and (45) respectively. The residual entropy in Eq. (52) is therefore consistent with that in the case $\Delta = -\infty$ in zero field shown in Sec. III.A.2, also with that of square ice

$$S/k_B = \frac{3}{2}\ln\left(\frac{4}{3}\right) . \tag{55}$$

## IV. Conclusions

In this article, we have studied the residual entropy of a two-dimensional Ising model with crossing and four-spin interactions. Following the work of Refs. 13 and 14, the exact solutions of the partition function in two soluble cases in zero field and one soluble case in an imaginary field $H_{ex} = i(\pi/2)k_B T$ are proposed by making use of the equivalence with the eight-vertex model. The residual entropy is then determined by the low temperature limit of the partition function and the ground state energy. In the free-fermion case in zero field, the ground states at zero temperature include the configurations disobeying the ice rules, which leads to a larger residual entropy than that of square ice. We have demonstrated, the residual entropy in this case is equal to

$$\frac{1}{8\pi^2}\int_0^{2\pi}d\theta\int_0^{2\pi}d\phi\ln\left[9-2\sqrt{2}\left(\cos\theta+\cos\phi\right)\right]\simeq 1.07052 \tag{56}$$

with the contribution from the configurations without (C). In another soluble case in zero field that $\Delta=-\infty$, we have modified the energy levels and defined the effective Hamiltonian and partition function. The ground states of the modified system are exactly the configurations obeying the ice rules. Therefore the residual entropy exactly agrees with that of square ice [Eq. (1)]. Finally, in the case that $\Delta=0$ in an imaginary field $H_{ex}=i(\pi/2)k_B T$, the partition function in the low temperature limit has the same behaviour as that in the case $\Delta=-\infty$ in zero field. Therefore the residual entropies in these two cases are equivalent. The solutions of these two cases can be seen as alternative approaches to the residual entropy problem of square ice.

Instead of using the transfer matrix method or the combinational methods, we have examined the extensive ground state degeneracy and the residual entropy of a two-dimensional system in a thermodynamic point of view. This work provides new insights into the exact evaluation of the residual entropy in frustrated systems, especially the exact solution of the two-dimensional ice, which is a famous classical problem in statistical physics. The method of taking the low temperature limit of the system can be used in future studies, e.g., the calculation of the residual entropy of three-dimensional frustrated spin systems and ice[40].

## Acknowledgements

This work was supported by Guangdong Basic and Applied Basic Research Foundation (Grant No. 2021A1515010328) and Key-Area Research and Development Program of Guangdong Province (Grant No. 2020B010183001).

## Statements and Declarations

The authors have no competing interests to declare that are relevant to the content of this article.